\documentclass[twocolumn,showpacs,superscriptaddress]{revtex4}

\usepackage{bm}         
\usepackage{amssymb}
\usepackage{amsmath}
\usepackage{graphics}
\usepackage{epsfig}
\usepackage[usenames]{color}

\begin{document}

\title{Superconducting quantum interference devices with submicron Nb/HfTi/Nb junctions for investigation
of small magnetic particles}

\author{J.~Nagel}
\affiliation{%
  Physikalisches Institut -- Experimentalphysik II and Center for Collective Quantum Phenomena in LISA$^+$,
  Universit\"at T\"ubingen,
  Auf der Morgenstelle 14,
  D-72076 T\"ubingen, Germany
}
\author{O.F.~Kieler}
\affiliation{%
  Fachbereich 2.4 "Quantenelektronik", Physikalisch-Technische Bundesanstalt,
  Bundesallee 100,
  38116 Braunschweig, Germany
}
\author{T.~Weimann}
\affiliation{%
  Fachbereich 2.4 "Quantenelektronik", Physikalisch-Technische Bundesanstalt,
  Bundesallee 100,
  38116 Braunschweig, Germany
}
\author{R.~W\"{o}lbing}
\affiliation{%
  Physikalisches Institut -- Experimentalphysik II and Center for Collective Quantum Phenomena in LISA$^+$,
  Universit\"at T\"ubingen,
  Auf der Morgenstelle 14,
  D-72076 T\"ubingen, Germany
}
\author{J.~Kohlmann}
\affiliation{%
  Fachbereich 2.4 "Quantenelektronik", Physikalisch-Technische Bundesanstalt,
  Bundesallee 100,
  38116 Braunschweig, Germany
}
\author{A.B.~Zorin}
\affiliation{%
  Fachbereich 2.4 "Quantenelektronik", Physikalisch-Technische Bundesanstalt,
  Bundesallee 100,
  38116 Braunschweig, Germany
}
\author{R.~Kleiner}
\affiliation{%
  Physikalisches Institut -- Experimentalphysik II and Center for Collective Quantum Phenomena in LISA$^+$,
  Universit\"at T\"ubingen,
  Auf der Morgenstelle 14,
  D-72076 T\"ubingen, Germany
}
\author{D.~Koelle}
\affiliation{%
  Physikalisches Institut -- Experimentalphysik II and Center for Collective Quantum Phenomena in LISA$^+$,
  Universit\"at T\"ubingen,
  Auf der Morgenstelle 14,
  D-72076 T\"ubingen, Germany
}
\author{M.~Kemmler}
\email{matthias.kemmler@uni-tuebingen.de}
\affiliation{%
  Physikalisches Institut -- Experimentalphysik II and Center for Collective Quantum Phenomena in LISA$^+$,
  Universit\"at T\"ubingen,
  Auf der Morgenstelle 14,
  D-72076 T\"ubingen, Germany
}

\date{\today}

\begin{abstract} 
We investigated, at temperature $4.2\,\mathrm{K}$, electric transport, flux noise and resulting spin sensitivity of miniaturized Nb direct current superconducting quantum interference devices (SQUIDs) based on submicron Josephson junctions with HfTi barriers.
The SQUIDs are either of the magnetometer-type or gradiometric in layout.
In the white noise regime, for the best magnetometer we obtain a flux noise $S_{\Phi}^{1/2}=250\,\mathrm{n}\Phi_0/\mathrm{Hz}^{1/2}$, corresponding to a spin sensitivity $S^{1/2}_\mu\,\ge\,29\,\mu_B/\mathrm{Hz}^{1/2}$.
For the gradiometer we find $S_{\Phi}^{1/2}=300\,\mathrm{n}\Phi_0/\mathrm{Hz}^{1/2}$ and $S^{1/2}_\mu\,\ge\,44\,\mu_B/\mathrm{Hz}^{1/2}$.
The devices can still be optimized with respect to flux noise and coupling between a magnetic particle and the SQUID, leaving room for further improvement towards single spin resolution.
\end{abstract}


\maketitle

\section{Introduction}

Growing interest in the investigation of small spin systems like molecular magnets\cite{Wernsdorfer01,Gatteschi03,Bogani08}, single electrons \cite{Bushev11} or cold atom clouds\cite{Fortagh05}, demands for proper detection schemes.
Compared to, e.g., magnetic resonance force microscopy \cite{Rugar04} or magneto-optic spin detection \cite{Maze08,Balasubramanian08}, superconducting quantum interference devices (SQUIDs) offer the advantage of direct measurement of changes of the magnetization in small spin systems \cite{Wernsdorfer01,Gallop03}.
High spin sensitivity requires SQUIDs with low flux noise and strong magnetic coupling between particle(s) and SQUID loop.
These needs can be met by nano-scaling the devices \cite{Foley09,Bouchiat09,Nagel11}, e.g., by focused ion beam milling \cite{Troeman07,Hao08}, electron-beam lithography \cite{Voss80a}, atomic force microscopy anodization \cite{Bouchiat01,Faucher09}, shadow evaporation \cite{Finkler10} or by coupling small pickup loops to larger SQUIDs \cite{Koshnick08}.
While nanopatterning of the SQUID loop yields no basic technical difficulties, the creation of overdamped Josephson junctions (JJs), as required for direct current (dc) SQUIDs, with submicron dimensions is more challenging.
A widely used approach is to use constriction JJs.
In some cases this yielded dc SQUIDs \cite{Voss80a,Hao08} with root mean square (rms) flux noise $S_{\Phi}^{1/2}$ down to $0.2\,\mu\Phi_0/\mathrm{Hz}^{1/2}$ ($\Phi_0$ is the magnetic flux quantum), which however are suitable only for operation in a limited range of temperature $T$.
Even smaller $S_{\Phi}^{1/2}=17\,\mathrm{n}\Phi_0/\mathrm{Hz}^{1/2}$ has been reported for larger SQUIDs based on superconductor-insulator-superconductor (SIS) tunnel JJs with external resistive shunts \cite{VanHarlingen82}.
In this letter, we report on the realization of small and sensitive dc SQUIDs based on S-normalconductor (N)-S sandwich-type JJs, without resistive shunts, which simplifies SQUID miniaturization.

sec\section{Sample Fabrication and layout}

\begin{figure}[b]
\includegraphics[width=8.5cm]{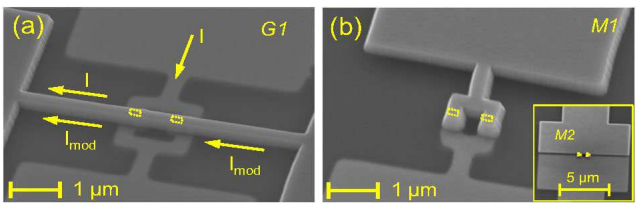}
\caption{(Color online) Scanning electron microscopy (SEM) images of the SQUIDs.
The JJs with size $200\times200\,\mathrm{nm}^2$ are indicated as dotted lines in the top Nb layer.
(a) Gradiometer {\it G1} with line width $250\,$nm and outer loop size $1.5\times1.5\,\mathrm{\mu m}^2$; arrows indicate scheme of current flow;
(b) Magnetometer {\it M1} with line width $250\,$nm and SQUID hole $500\times500\,\mathrm{nm}^2$.
Inset: washer-type magnetometer {\it M2} with washer area $10\times10\,\mathrm{\mu m}^2$ and SQUID hole $500\times500\,\mathrm{nm}^2$.
\label{Fig:SQUID_layouts}}
\end{figure}

Our JJs are based on a Nb/HfTi/Nb trilayer process \cite{Hagedorn02}, which was developed for the fabrication of submicron SNS junctions \cite{Hagedorn06}.
All JJs are square shaped with lateral dimensions $200\times200\,\mathrm{nm^2}$.
The JJs with barrier thickness $d_{\mathrm{HfTi}}$ = $24\,\mathrm{nm}$ have a critical current density $j_c \approx  200-300\,\mathrm{kA/cm}^2$ at $T=4.2\,$K and a resistance times junction area $\rho_n\,\approx\,14-19\,\mathrm{m}\Omega\mathrm{\mu m^2}$, leading to a characteristic voltage $V_c=j_c\rho_n\approx40\,\mu\mathrm{V}$.
The three SQUIDs presented in this paper have different layouts.
{\it G1} [see Fig.~\ref{Fig:SQUID_layouts}(a)] has a gradiometric design.
The gradiometer line in the top Nb layer carries the bias current $I$ (flowing through the junctions to the bottom Nb layer) and in addition allows for the (on-chip) application of magnetic flux $\Phi$ to the gradiometer (referred to one loop) via a current $I_\mathrm{mod}$ without the need of external coils.
{\it M1} [see Fig.~\ref{Fig:SQUID_layouts}(b)] is of the magnetometer-type.
{\it M2} [see inset of Fig.~\ref{Fig:SQUID_layouts}(b)], which is similar to {\it M1}, has a washer, allowing flux modulation with relatively small external magnetic fields ($B/\Phi = 0.5\,\mathrm{mT/\Phi_0}$).

sec\section{Experiments}

All measurements were performed at $T=4.2\,\mathrm{K}$ in a high-frequency shielding chamber with the sample mounted inside a magnetic shield.
All currents were applied by battery powered low-noise current sources.
For the noise measurements we used a commercial Nb dc SQUID amplifier surrounded by a superconducting Nb shield \cite{Starcryo}.
The SQUID is connected in parallel to the input coil of the SQUID amplifier, with an input resistor $R_{\mathrm{in}}$ connected in series with the coil.
A separate feedback (and modulation) coil of the SQUID amplifier allows for a flux locked loop operation of the SQUID amplifier with a sensitivity $S_{V,\mathrm{amp}}^{1/2} \approx 40\,\mathrm{pV}/\mathrm{Hz}^{1/2}$ for $R_{\mathrm{in}}=3.3\,\Omega$ at $T=4.2\,\mathrm{K}$.
The typical bandwidth of the amplifier is of the order of few tens of kHz.
To determine the rms flux noise of our SQUIDs we measured the voltage noise at the output of the amplifier.
After subtracting the noise contribution from the amplifier, we obtain the spectral density of voltage noise $S_{\mathrm{V,SQUID}}$ for the SQUID and calculate the corresponding rms flux noise $S_\Phi^{1/2}=S_{\mathrm{V,SQUID}}^{1/2}/|\partial V/\partial \Phi|$.
Here, $V$ is the voltage across the SQUID and $\partial V/\partial \Phi$ is the transfer coefficient.

\begin{figure}[bt]
\includegraphics{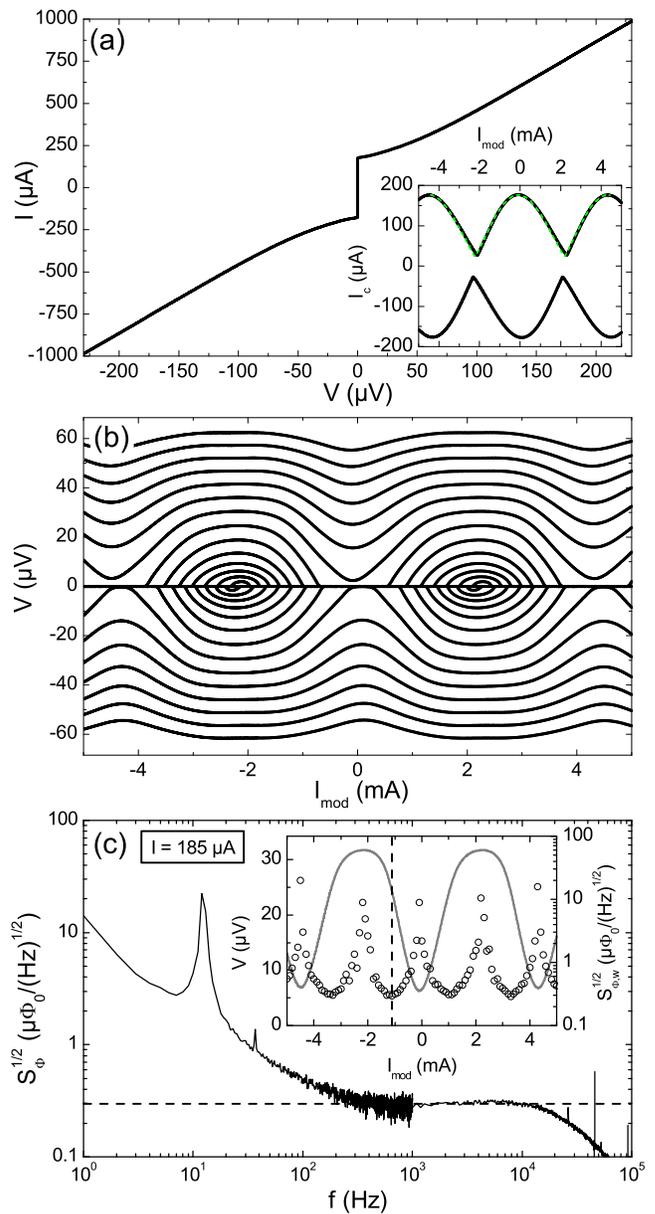}
\caption{(Color online) Transport and noise characteristics of {\it G1} at $\mathrm{T}=4.2\,K$:
(a) IVC at $I_\mathrm{mod}=0$; inset shows measured $I_c(I_\mathrm{mod})$ (solid line) and simulated curve (dashed line).
(b) $V(I_\mathrm{mod})$ for $I=-297\ldots300\,\mu$A (in $20.1\,\mu$A steps).
(c) Spectral density of rms flux noise $S_{\Phi}^{1/2}(f)$ at optimal working point (c.f.~dashed line in inset); dashed line indicates $300\,\mathrm{n}\Phi_0/{\rm Hz}^{1/2}$.
Inset: $V$($I_\mathrm{mod}$) (solid line) and $S^{1/2}_{\Phi,w}$($I_\mathrm{mod}$) (open circles; averaged from $f=$ 2 to 3\,kHz).
}
\label{Fig:SQUID_low_field}
\end{figure}

Figure \ref{Fig:SQUID_low_field}(a) shows the current voltage characteristic (IVC) of {\it G1} measured at $I_\mathrm{mod}=0$.
The IVC is resistively shunted junction (RSJ)-like, with a critical current $I_c=178\,\mu\mathrm{A}$ and resistance $R=233\,\mathrm{m}\Omega$, yielding $V_c=41.5\,\mu\mathrm{V}$.
The inset of Fig.~\ref{Fig:SQUID_low_field}(a) shows $I_c(I_\mathrm{mod})$ together with a simulated curve based on the RSJ model (including thermal noise and inductance asymmetry), which yields $\beta_L\equiv 2I_0L/\Phi_0=0.18$.
Here, $I_0$ is the average maximum critical current of the two JJs, and $L$ is the inductance of the gradiometric SQUID, i.e.~half the inductance of one loop of the gradiometer.
With $2I_0=178\,\mathrm{\mu A}$ we obtain $L=2.1\,\mathrm{pH}$.
From the measured period of $I_c(I_\mathrm{mod})$ we obtain $\Phi/I_\mathrm{mod}=227\,\mathrm{m}\Phi_0/\mathrm{mA}$.
The small but finite shift $\Delta I_\mathrm{mod}=95\,\mu\mathrm{A}$ of the maxima in $I_c(I_\mathrm{mod})$ for opposite polarity can be solely attributed to an inductance asymmetry due to the asymmetric current bias, i.e.~the asymmetry in the critical currents of the JJs is negligibly small.
Figure \ref{Fig:SQUID_low_field}(b) shows $V(I_\mathrm{mod})$ for different values of $I$.
For $I \approx 185\mu\mathrm{A}$ we obtain a maximum transfer coefficient $V_\Phi\approx 100\,\mu\mathrm{V}/\Phi_0$.
The inset of Fig.~\ref{Fig:SQUID_low_field}(c) shows $V(I_\mathrm{mod})$ and $S^{1/2}_{\Phi,\mathrm{w}}(I_\mathrm{mod})$ in the white noise regime (determined by averaging the spectra from $f=2$ to $3\,\mathrm{kHz}$) for $I=185\mu\mathrm{A}$.
This yields minima in $S^{1/2}_{\Phi,\mathrm{w}}(I_\mathrm{mod})$ at the optimum flux bias point (indicated by the dashed line), for which the main graph of Fig.~\ref{Fig:SQUID_low_field}(c) shows $S_{\Phi}^{1/2}$ vs frequency $f$.
For low frequencies $f\le 10\,\mathrm{Hz}$ we find $S_{\Phi}(f)\propto 1/f^2$, which can be attributed to a single fluctuator (flux or $I_c$) producing random telegraph noise in the time trace $V(t)$.
For higher frequencies $10\,\mathrm{Hz}\le f \le1\,\mathrm{kHz}$ the frequency dependence is more 1/$f$ like, which might be caused by an admixture of noise from a few additional fluctuators with higher characteristic frequencies.
The peak in $S_{\Phi}(f)$ near $f=12\,\mathrm{Hz}$ presumably results from mechanical vibrations.
The spectrum in the white noise limit above $1\,\mathrm{kHz}$ yields $S^{1/2}_{\Phi,\mathrm{w}}\,\approx\,300\,\mathrm{n}\Phi_0/{\rm Hz}^{1/2}$, with a cutoff
at $f\approx 2\times 10^4\,\mathrm{Hz}$ due to the SQUID amplifier electronics.
The magnetometer-type devices {\it M1} and {\it M2} had similar characteristics, with $S^{1/2}_{\Phi,\mathrm{w}}\,\approx\,250\,\mathrm{n}\Phi_0/{\rm Hz}^{1/2}$ and $\approx 270\,\mathrm{n}\Phi_0/{\rm Hz}^{1/2}$, respectively.

sec\subsection{estimated Spin sensitivity}

\begin{figure}[ht]
\includegraphics[width=8.49cm]{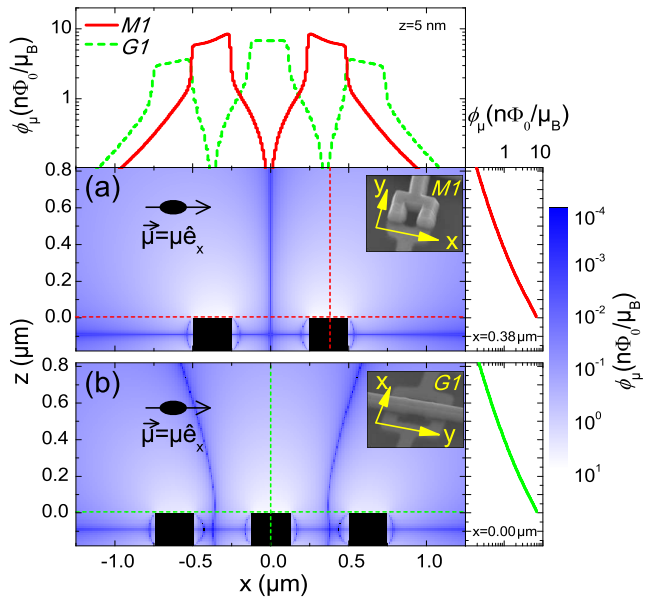}
\caption{(Color online) Calculated coupling factor $\phi_\mu$ vs particle position.
Main graphs show contour plots $\phi_\mu(x,z)$ for (a) magnetometer {\it M1} and (b) gradiometer {\it G1}; Nb structures are indicated by black rectangles.
Insets show SEM images of the SQUIDs.
Dashed lines indicate position of linescans $\phi_\mu(x)$ [shown above (a)] and $\phi_\mu(z)$ [shown to the right of (a) and (b)].}
\label{Fig:coupling_factor}
\end{figure}

Finally, we turn to the spin sensitivity $S^{1/2}_\mu = S^{1/2}_\Phi/\phi_\mu$ of our devices which, besides the flux noise, depends on the coupling factor $\phi_\mu$, i.e.~the amount of flux coupled into the SQUID by a magnetic particle, divided by the modulus $|\vec{\mu}|$ of its magnetic moment.
Taking into account the SQUID geometry, Fig.~\ref{Fig:coupling_factor} shows the calculated coupling factor of {\it M1} and {\it G1} vs.~the position $\vec{r}$ of a point-like magnetic particle with its magnetic moment $\vec{\mu}$ pointing in-plane of the SQUID loop.
A detailed description of the calculation procedure for non-gradiometric SQUIDs can be found in Ref.~[\onlinecite{Nagel11}].
For the gradiometric SQUID {\it G1} one has to consider the magnetic field distribution $\vec{B}(\vec{r})$ created by two circular currents $I_{1,2}=\pm I_B$ in each loop.
In this case the coupling factor $\phi_\mu$ is given by $\vec{B}(\vec{r})/2I_B$.
For an in-plane magnetization of the particle, layout {\it M1} provides the highest coupling factor if the particle is placed directly on top of the SQUID loop.
For {\it G1} the optimum coupling can be achieved if the particle is placed on the center conductor line.
At this position the particle couples flux of opposite sign into both loops of the gradiometric SQUID, which leads to an approximately twice as large coupling factor as compared to placing the particle on the outer conductors.
For a particle with 10\,nm diameter, placed directly on top of the SQUID, we take a vertical distance $z=5\,$nm from the SQUID surface, which yields $\phi_\mu=8.5\,\mathrm{n}\Phi_0/\mu_B$ ($\mu_B$ is the Bohr magneton) for {\it M1} and $6.8\,\mathrm{n}\Phi_0/\mu_B$ for {\it G1} at the center conductor.
With $S^{1/2}_{\Phi}\approx 250\,\mathrm{n}\Phi_0/\mathrm{Hz}^{1/2}$ we calculate the spin sensitivity of {\it M1} to $S^{1/2}_\mu = 29\,\mu_B/\mathrm{Hz}^{1/2}$.
For the gradiometric SQUID we calculate $S^{1/2}_\mu = 44\,\mu_B/\mathrm{Hz}^{1/2}$.

sec\section{Conclusions}

In conclusion, we have shown that miniaturized dc SQUIDs based on sandwich-type overdamped SNS Josephson junctions have a compact design and can be operated with very promising values of flux noise and spin sensitivity.
Although our devices are not optimized yet, flux noise values down to $250\,\mathrm{n}\Phi_0/\mathrm{Hz}^{1/2}$ have been achieved, leading to an estimated spin sensitivity as low as $29\mu_B/\mathrm{Hz}^{1/2}$.
Further improvements are feasible; e.g., placing the two SQUID arms on top of each other, as in Ref.[\onlinecite{VanHarlingen82}], allows for reduction of the SQUID inductance and hence of the flux noise.
Furthermore, the coupling can be improved by patterning an additional constriction within the SQUID loop.

sec\section*{Acknowledgment}

This work was supported by the Nachwuchs\-wissenschaftlerprogramm of the Universit\"{a}t T{\"u}bingen, by the Deutsche Forschungsgemeinschaft (DFG) via the SFB TRR21 and by the European Research Council via SOCATHES.
J.~Nagel and M.~Kemmler gratefully acknowledge support by the Carl-Zeiss Stiftung.

\end{document}